\documentstyle[epsfig,12pt]{article}
\begin{document}
\begin{center}
{ON QUARK-LEPTON SYMMETRY}
\end{center}
\vspace{3mm}
\begin{center}
{N.A. Korkhmazyan, N.N. Korkhmazyan\\
5 Khanjyan St., Armenian Pedagogical Institute, Yerevan 375010, Armenia\\
E-mail: armped@moon.yerphi.am}
\end{center}
\vspace{5mm}
\begin{center}
{Abstract}
\end{center}

The new quantum number   is introduced. It is shown that the conservation of 
-number results in the conservation of difference between baryon and 
lepton numbers. The problem of quark-lepton symmetry is discussed.  It 
is shown that the nature of quark-lepton symmetry stems from the fact 
that the particles of one generation are subject to the symmetry 
transformation represented by 4-group of diedr.vspace{5cm}\\

The problem of quark-lepton symmetry \cite{1}-\cite{2} is closely 
related to the additive quantum number $\sigma$  introduced in the work 
\cite{3} with the aim to bring in some symmetry to the asymmetric 
disposition of upper and lower quarks along the "charge axis". The newly 
introduced quantum number is determined so as to result, in combination 
with quark electric charge q, the charge of the respective lepton.
Then for $u$  and $d$  quarks we will have, respectively, 
$\sigma_u=1/3$  and $\sigma_d=-2/3$. In general,$\sigma$ -numbers for 
all quarks (and antiquarks) are determined by the following formula:
\begin{equation}
\label{1}
\sigma=q-1/3, \quad \tilde{\sigma}=\tilde{q}+1/3
\end{equation}

Now the quarks and leptons are positioned symmetrically along 
$\sigma=-q$  axis on $(q,\sigma)$ plane (Fig 1.).
 
\epsfig{file=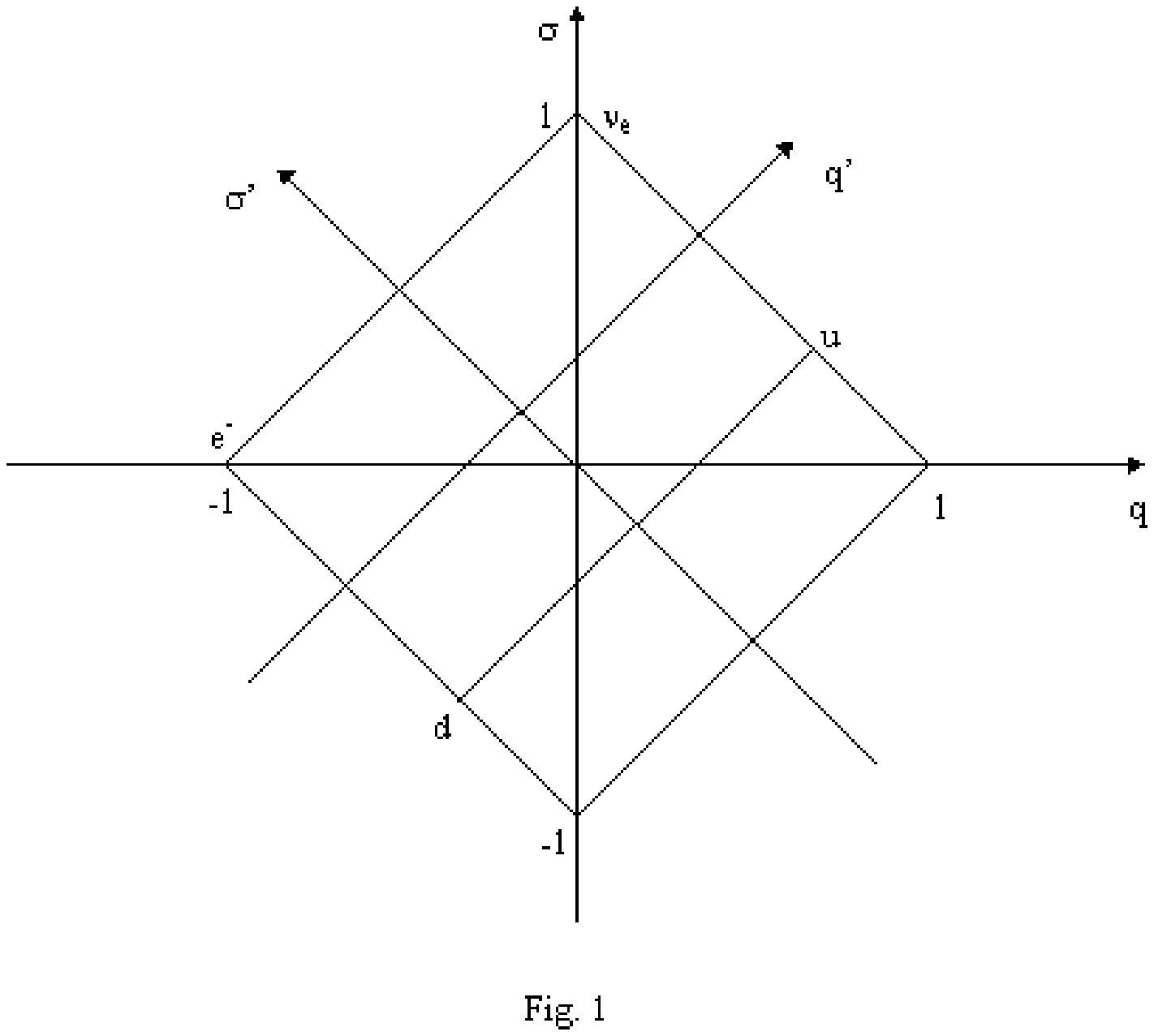,width=17cm,height=17cm}

This brings about an idea that together with the law of conservation of 
electric charge there is realized the second law of  conservation, the 
law of conservation of $sigma$ -number. As we will see later, this idea 
results in extremely valuable findings. I

In particular, we will show that the conservation of $sigma$ - number 
results in conservation of the difference between baryon and lepton 
numbers obtained earlier by Georgi and Glashow from SU(5) symmetry \cite{4}.
Let us first show that $\sigma$ - numbers for baryons, mesons, leptons 
and  $\gamma$-photons are determined by the following equations:
\begin{equation}
\label{2}
\sigma_B=Q_B-1,\quad \sigma_M=Q_M, \quad \sigma_L=Q_L+1,\quad 
\sigma_\gamma=0, \quad \tilde{\sigma}=-\sigma, \end{equation}
where $Q$ is an electric charge of a particle and the sign ~ denotes 
anti. Indeed, equations (\ref{2}) are obvious for baryons and mesons due 
to their structure. They are also applicable to exotic baryons and 
mesons. Using (\ref{2}), we can obtain, in particular,$\sigma(p)=0,\quad 
\sigma(n^0)=-1, \quad \sigma(\pi^0)=0,\quad 
11~\sigma(\tilde{\wedge^0})=1,\quad \sigma(\sum^-)=-2$, and so on.
Taking into account that $\sigma (\pi^0)=0$, we obtain  $\sigma_\gamma =0$ 
from $\pi^0 \Rightarrow 2\gamma$ decay. In order to find $\sigma_l$ , 
we will give the same quantum number to all charged leptons 
$(\sigma_{e-}=\sigma_{\mu-}=\sigma_{\tau-})$ and 
another quantum number to neutral leptons 
$(\sigma_{\nu_e}=\sigma_{\nu_\mu}=\sigma_{\nu_\tau})$. These numbers are 
related to each other by the formula $\sigma_{\nu_e}=1+\sigma_e$   
derived from $n^0 \Rightarrow p+e^-+\tilde{\nu}$ reaction.
Then from $uu \Leftarrow x \Rightarrow e+\tilde{d}$ reaction where the 
same boson may decay into antilepton + 
antiquark or quark pair we derive $\sigma_{e^+}=-\sigma_{e^-}=0$, and 
$\sigma_\nu=1$. As we can see, these results 
are in accordance with equations (\ref{2}). We can rewrite them as follows:
\begin{equation}
\label{3}
\sum{Q_B} - N_B + \sum{\tilde{Q}_B} + \tilde{N}_B + \sum{Q_M} + 
\sum{\tilde{Q}_m} + \sum{Q_L} + N_L + \sum{\tilde{Q}_L} -\tilde{N}_L=const,
\end{equation}
where $N$  is the number of particles. Here, taking into account the 
conservation of electric charge, we will finally obtain the desired result:
\begin{equation}
\label{4}
(N_B - \tilde{N}_B) - (N_L - \tilde{N}_L)=const
\end{equation}
 
Now we will focus on the phenomenon of quark - lepton symmetry. Let us 
shift to the new system of coordinates $(q',\sigma'$) with the point of 
origin in the center of a rectangle $e\nu_eud$ and axes directed along the 
axes of symmetry (see Fig1.). The relation with an original
system of coordinates is given by the following expressions:
\begin{equation}
\label{5}
q'=\frac{1}{\sqrt{2}}q+\frac{1}{\sqrt{2}}\sigma;\quad 
\sigma'=-\frac{1}{\sqrt{2}}q+\frac{1}{\sqrt{2}}\sigma-\frac{1}{3\sqrt{2}}
\end{equation}

The coordinates of the vertices are now as follows:
\begin{equation}
\label{6}
e\left(-\frac{1}{\sqrt{2}},\frac{\sqrt{2}}{3}\right);\quad
\nu_e\left(\frac{1}{\sqrt{2}},\frac{\sqrt{2}}{3}\right);\quad
u\left(\frac{1}{\sqrt{2}},-\frac{\sqrt{2}}{3}\right);\quad
d\left(-\frac{1}{\sqrt{2}},-\frac{\sqrt{2}}{3}\right)
\end{equation}

The group that represents the symmetry transformation for the above mentioned 
rectangle, consists of the following elements:\\
$E$ - identity transformation; turn around axis $z$ for angle $2\pi$;\\
$A$ - turn around axis $\sigma'$ for angle $\pi$;\\
$B$ - turn around axis $q'$  for angle $\pi$;\\
$C$ - turn around axis z for angle $\pi$.

For these elements we will obtain the group multiplication table (see Table 1).
\begin{center}
{\bf{Table 1}}\vspace{1.5cm}\\
\begin{tabular}{|c|cccc|}
\hline
  & E & A & B & C\\ \hline
E & E & A & B & C\\
A & A & E & C & B\\
B & B & C & E & A\\
C & C & B & A & E\\
\hline
\end{tabular}
\end{center}
\vspace{1cm}

Under multiplication we understand the subsequent execution of the 
corresponding operations. Besides that, all the elements of the group have 
order 2 (except for identity element $E$), since $\chi^2=E$ and 
$\chi^{-1}=\chi$, where $\chi$  is an arbitrary element of the group. Thus, 
the totality of the elements $E,A,B,C$ makes up the Abelian group. The 
similitude transformation of a regular polygon is expressed by means of the 
following matrices \cite{5}
\begin{equation}
\label{7}
{D_k}=
\left(\begin{array}{cc}
\cos{\frac{2\pi k}{n}}, & \sin{\frac{2\pi k}{n}}\\
-\sin{\frac{2\pi k}{n}}, & \cos{\frac{2\pi k}{n}}
\end{array}\right) \quad
{U_k}=
\left(\begin{array}{cc}
-\cos{\frac{2\pi k}{n}}, & \sin{\frac{2\pi k}{n}}\\
\sin{\frac{2\pi k}{n}}, & \cos{\frac{2\pi k}{n}}
\end{array}\right)
\end{equation}
where k=0,1,2,…, n-1. These 2n-dimensional matrices make up the group oforder 
2n known as the group of diedr. In case of n=2 we have the simplest 
case of the group with elements
\begin{equation}
\label{8}
{E}=
\left(\begin{array}{cc}
1 & 0\\
0 & 1
\end{array}\right);
\quad
{A}=
\left(\begin{array}{cc}
-1 & 0\\
0 & -1
\end{array}\right);
\quad
{B}=
\left(\begin{array}{cc}
-1 & 0\\
0 & 1
\end{array}\right);
\quad
{C}=
\left(\begin{array}{cc}
1 & 0\\
0 & -1
\end{array}\right);
\end{equation}

Taking into account that the group (\ref{8}) is isomorphic to the above 
mentioned group, we come to conclusion that (\ref{8}) is the matrix 
representation of the group. Thus, if particles of the same generation 
are located in the vertices of rectangle (6), on the pla
ne ($q',\sigma'$), this distribution is subject to the symmetry 
transformation 
described by the group (\ref{8}). It is also easy to show that if 
distribution 
of particles on the plane ($q',\sigma'$) is subject to the symmetry 
transformation described by the group (\ref{8}) and if coordina
tes of any arbitrary particle from (\ref{6}) coincide with one of the 
vertices, there should be three more particles (and only three, without taking
into account the color of the quarks) whose coordinates coincide with 
the remaining vertices.

In conclusion, we can say that the nature of the quark-lepton symmetry could 
be explained by  -symmetry, which, in its turn, is caused by newly introduced 
in the work \cite{1} quantum number $\sigma$ . In other words, the 
quark-lepton symmetry stems from the fact tha
t particles of the same generation are subject to the symmetry transformation 
represented by 4-group of diedr.

We wish to thank Prof. Andrey Amatuni for comments and valuable suggestions.


\begin{thebibliography}{99}
\bibitem{1} L.B.Okun, Physics of elementary particles, (Moscow, 1987).
\bibitem{2} F.Halzen, A.Martin, Quarks and leptons, (New York, Chichester, 
Brisbane, Toronto, Singapore, 1984).
\bibitem{3} N.A. Korkhmazyan, Docl. NAN Armenii, Vol. 99 (1999) 182.
\bibitem{4} H.Georgi, Sh.Glashow, Phys. Rev. Lett., Vol. 32 (1974) 438.
\bibitem{5} E.Wigner, Group Theory, (New York and London, 1959).
\end{thebibliography}
\end{document}